\documentclass[twocolumn,aps,prl,nofootinbib]{revtex4-2} 
\usepackage{amsfonts,amsmath,amssymb,mathrsfs}
\usepackage{mathptmx,mathtools,array,latexsym}
\usepackage{xcolor}
\definecolor{patriarch}{RGB}{139,69,19}
\usepackage[colorlinks=true,linkcolor=blue,citecolor=blue,urlcolor=blue]{hyperref}

\def\be{\begin{equation}}
\def\ee{\end{equation}}
\def\bs{\begin{split}}
\def\es{\end{split}}
\def\bea{\begin{eqnarray}}
\def\eea{\end{eqnarray}}
\def\ba{\begin{aligned}}
\def\ea{\end{aligned}}
\def\nn{\nonumber}
\def\p{\partial}

\def\cN{\mathcal{N}}

\def\cT{\mathcal{T}}

\begin{document}

\title{CFT-Inspired Two-Horizon Thermodynamics as a Diagnostic for the Bi-Hair
Organization of NUT Charge}

\author{Di Wu$^{1,2}$}
\email{Contact author: wdcwnu@163.com, d277wu@uwaterloo.ca}

\author{Robert B. Mann$^2$}
\email{Contact author: rbmann@uwaterloo.ca}

\author{Shuang-Qing Wu$^1$}
\email{Contact author: sqwu@cwnu.edu.cn}

\affiliation{$^1$School of Physics and Astronomy, China West Normal University,
Nanchong, Sichuan 637002, People's Republic of China \\
$^2$Department of Physics and Astronomy, University of Waterloo, Waterloo, Ontario N2L 3G1, Canada}

\date{\today}

\begin{abstract}
After decades without a fully self-consistent thermodynamic formulation, recent proposals for
Lorentzian Taub-NUT spacetime have yielded several inequivalent descriptions, each satisfying its
own first law and Smarr relation, leaving unresolved which formulation is physically preferred.
In this Letter, we introduce a two-horizon diagnostic whose left- and right-moving
thermodynamic temperatures are fixed via the harmonic mean relations before the thermodynamic
variables are chosen. For uncharged Taub-NUT spacetime, the left-moving first law
closes with the mass and NUT charge, whereas, within the restricted homogeneous class examined
here, the right-moving first law requires an additional thermodynamic secondary
hair $J_n = mn$. This result supports a bi-hair thermodynamic organization of the NUT parameter
through $N = n$ and $J_n = mn$. Furthermore, this two-horizon framework offers a new opportunity
toward addressing similar thermodynamic non-uniqueness in other two-horizon NUT-charged systems
when the first law and Smarr relation alone do not identify a physically preferred formulation.
\end{abstract}

\maketitle

\textit{Introduction}---Black holes have become observational laboratories for strong field
gravity through gravitational wave observations of binary mergers by ground-based interferometers
\cite{PRL116-061102,PRL136-041403}, and horizon-scale imaging by the Event Horizon Telescope
\cite{APJL875-L1,APJL930-L12}. Gravitational wave observations have also enabled direct tests
of Hawking's area law, from GW150914 to the more recent GW250114 event \cite{PRL127-011103,
PRL135-111403}, bringing a foundational aspect of black hole thermodynamics into increasingly
close contact with observation. These developments sharpen the need for a well-defined thermodynamic
description of spacetimes with nontrivial horizon structures. In this context, the Taub-NUT spacetime
is an exact stationary, axisymmetric vacuum solution of Einstein's equations, in which the NUT
parameter characterizes a gravo-magnetic monopole \cite{RMP70-427}. Like the Kerr black hole, it
possesses a two-horizon structure, and its analytic tractability makes it a useful model that may
shed new light on astrophysical black hole thermodynamics.

However, more than seventy years after the discovery of the Taub-NUT solution \cite{AM53-472,JMP4-915}
and more than half a century after the birth of black hole thermodynamics \cite{PRD7-2333,CMP31-161},
there is still no consensus on the physically preferred thermodynamic description of the NUT-charged
spacetimes. This multiplicity contrasts with the more familiar thermodynamic descriptions of standard black holes \cite{CMP31-161}.
Several inequivalent thermodynamic formulations have been proposed, including the Misner charge
formulation, modified- and horizon-mass approaches, and bi-hair formalism, etc, each satisfying
its own macroscopic first law and Smarr relation \cite{PRD100-101501,PRD105-124013,PLB846-138227,
2606.03958,PRD108-064034,PRD108-064035,PRD100-064055,CQG36-194001,PLB798-134972,JHEP0520084,EPJC83-365,
EPJC83-589,PRD101-124011,PRD105-124034,JHEP1022174,IJMPD31-2250021}. The traditional thermodynamic
criteria therefore cannot determine which formulation is physically more meaningful by themselves.
Together, the intrinsic structure with two horizons and the macroscopic nonuniqueness make the Taub-NUT
a controlled setting for asking whether thermodynamics of the two horizons can constrain the candidate
state spaces without first choosing one macroscopic first law convention.

A central distinction among these schemes concerns whether the NUT parameter is assigned
one or more thermodynamic response roles. While most of these formulations \cite{PRD100-064055,
CQG36-194001,PLB798-134972,JHEP0520084,PRD101-124011,PRD105-124034,JHEP1022174,IJMPD31-2250021} adhere
to the traditional framework of the usual black hole thermodynamics and retain the same number of independent thermodynamic variables as solution parameters, the bi-hair strategy
\cite{PRD100-101501,PRD105-124013} provides a new perspective that does not regard NUT charge only as a
variable but views it as a thermodynamic bi-hair in the enlarged thermodynamic state space, just like a coin with two facets. From this novel idea one can show  not only the Christodoulou-Ruffini-type squared-mass formula \cite{PRL25-1596,PRD4-3552} can be naturally extended to the whole family of  NUT-charged spacetimes, but also the universal area (entropy) products \cite{PRL106-121301} can be satisfactorily explained \cite{PLB807-135521} (vs \cite{IJMPA38-2350090}) within the bi-hair representation. If not solely limited to the first law and Smarr relation, these two features provide complementary support for the bi-hair scheme. Below we will provide further supporting evidence \cite{2609.33032} from the Kerr/CFT correspondence and hidden symmetry that favors our bi-hair formalism.

Rather than constructing another thermodynamically consistent description for horizon thermodynamics,
here we develop a two-horizon diagnostic based on thermodynamic quantities descended from both
horizons. Motivated by the left- and right-moving thermodynamics compatible with the black
hole/CFT studies \cite{PRD56-4994,PLB608-251,JHEP0909088,JHEP1112017,JHEP0313102}, one can define
the left- and right-moving (LRM) entropies via the sum and difference of the outer and inner
horizon entropies,
\be
S_L = (S_+ +S_-)/2 \, ,\qquad S_R = (S_+ -S_-)/2 \, ,
\ee
which in related black hole/CFT settings correspond to the respective entropies of the left and right moving excited modes of effective string and D-brane models,
with different respective temperatures in a non-chiral two-dimensional conformal field theory. Their associated temperatures are determined via the harmonic mean relations:
\be
\frac{1}{T_L} = \frac{1}{T_+} +\frac{1}{T_-} \, ,\qquad
\frac{1}{T_R} = \frac{1}{T_+} -\frac{1}{T_-} \, ,
\label{hmr}
\ee
where the subscripts $\pm$ label the outer and inner horizons, and $T_-$ is taken to be negatively
signed, as in \cite{PLB608-251}. Here $T_L$ and $T_R$ are effective temperatures of the entropy sum and difference, distinct from the horizon Hawking temperatures $T_\pm$. Crucially, the left- and right-temperatures are fixed by Eq. (\ref{hmr}) before the thermodynamic variables are chosen. A candidate state space is then checked by requiring the same extensive variables to reproduce these fixed temperatures and simultaneously satisfy the LRM first laws and Smarr relations. The diagnostic therefore tests the thermodynamic state space rather than assuming it from the outset.

Applying this diagnostic to the uncharged Taub-NUT spacetime leads to a clear result. The
left-moving (LM) first law closes using only the mass and NUT charge, whereas, within
the restricted homogeneous class examined here, the right-moving (RM) first law requires
additionally the thermodynamic secondary hair
\be
J_n = mn \, ,
\ee
which appears in the next-to-leading term ($\cos\theta/r$) of $g_{t\phi}$ in the large $r$
asymptotic expansion;  see Eq. (7) of Ref. \cite{PRD105-124013} and Eqs. (A3-A4) of Ref.
\cite{PRD108-064035} for a formal computation. Also, it can be identified with the gravitational
mass of a five-dimensional Kaluza-Klein magnetic monopole \cite{PLB634-531}. Furthermore, it is
related to the gyromagnetic ratios in the Kerr-Taub-NUT-type spacetimes \cite{PRD77-044038,PRD77-124022},
gravitational actions of four-dimensional Taub-NUT-type spacetimes \cite{PRD72-084032,PRD75-024007},
and angular-momentum-like properties \cite{CQG22-3555,PLB750-591,CQG23-4473}. An alternative such as $X_\alpha=mn+\alpha n^2$ can be thermodynamically equivalent,
but the added term lacks the independent geometric motivation discussed above. In this sense,
$J_n=mn$ is the physically preferred representative, rather than a mathematically unique off-shell coordinate.

The precise sense is that the two-horizon diagnostic favors a bi-hair thermodynamic formalism
of the NUT parameter through $N = n$ and $J_n = mn$. Importantly, $J_n$ is defined off shell as a
thermodynamic response variable rather than as a new asymptotic charge or metric parameter. This
evidence is not meant to imply that the Misner string or the Euclidean regularity issue is irrelevant.
Rather, the diagnostic asks which thermodynamic variables are favored when the thermodynamic quantities
of the two horizons are required to admit the standard left/right decomposition.

\textit{Kerr benchmark}---It is useful to first recall how the left/right thermodynamic
decomposition works in an ordinary black hole, where the physical interpretation of the variables
is not controversial. In the Kerr case \cite{PLB608-251} with $J = Ma$, the entropies of its
two horizons ($r_{\pm} = M\pm \sqrt{M^2 -a^2}$) are: $S_{\pm} = \pi(r_{\pm}^2+a^2) = 2\pi\, Mr_{\pm}$.

The LM entropy is: $S_L = 2\pi\, M^2$, which is solely described by the energy $M/2$. Its
thermodynamic description is insensitive to the sign of the horizon separation and does not require
the angular momentum as an independent thermodynamic variable.

The RM entropy is: $S_R = 2\pi\sqrt{M^4 -J^2}$, which depends on the angular momentum $J = Ma$
in addition to the energy $M/2$, and therefore its RM first law cannot close in a state space
that contains only the mass. In other words, the LM first law closes with the energy $M/2$, whereas
the RM first law requires an extra $J$. Thus, in the familiar Kerr case, the left/right
decomposition separates a LM contribution that is blind to rotation from a RM contribution
that carries rotation. The Kerr example therefore motivates the question to be asked in the Taub-NUT
case: does the RM description reveal a thermodynamic response that is not visible in the LM
description?

\textit{Macroscopic thermodynamic schemes}---Several macroscopic schemes have been developed for
the Taub-NUT thermodynamics till now. Notably, the most popular Misner charge (`$\Psi-\cN$')
formulation \cite{PRD100-064055,CQG36-194001,PLB798-134972,JHEP0520084} assigns thermodynamic
variables to the Misner string contribution; modified mass \cite{PRD101-124011,PRD105-124034}
and horizon mass \cite{JHEP1022174} formulations alter the thermodynamic energy; the $J_n-\Xi$
construction \cite{IJMPD31-2250021} also introduces an extra variable $J_n = mn$. These
approaches clarify different aspects of the problem and can yield consistent macroscopic first
laws. Our purpose here is not to question their internal consistency, but to ask which variables
are favored by the additional LRM criterion.

This distinction is important because the NUT parameter mixes several geometric roles. It is a
gravito-magnetic charge and is tied to the Misner string. The LRM analysis below checks whether
the thermodynamic quantities of the two horizons expose a response channel whose scaling resembles
that of a rotational extensive variable. A macroscopic first law can emphasize one of these features
by a suitable choice of thermodynamic potential or energy. The LRM test is sensitive to a
different question: after the horizon entropies are reorganized into LRM combinations, which
variables are needed for the LRM first laws to be integrable and to satisfy the corresponding
Smarr formulae?

This perspective also sharpens the comparison with earlier formulations. The Misner charge, modified
mass, and horizon mass descriptions indicate that several choices of macroscopic variables can be made
compatible with a first law. That macroscopic compatibility, however, does not by itself decide which
state space carries the thermodynamic structure associated with the two horizons. The left/right
decomposition supplies an additional criterion. The calculation below asks whether the state space
based only on the mass and the NUT charge is sufficient once the same Taub-NUT thermodynamic quantities
are required to admit a consistent LRM thermodynamic description, or whether a minimal homogeneous
enlargement is favored.

For reference, the End Matter collects the candidate homogeneous embedding, a complementary
on-shell consistency check of the LRM first laws, and explicit representatives of the competing
macroscopic thermodynamic schemes. The ordinary horizon first law, squared-mass formula, and
entropy product serve only as admissibility checks; the discriminator in the main text is the
LRM closure test itself.

\textit{LRM thermodynamics constructed from two horizons}---We now apply the LRM criterion to
the Lorentzian Taub-NUT solution with the Misner strings symmetrically distributed on the axis,
whose metric is
\be\begin{aligned}
ds^2 = & -f(r)(dt +2n\cos\theta\, d\phi)^2 +dr^2/f(r) \nn \\
& +\big(r^2+n^2\big)\big(d\theta^2 +\sin^2\theta\, d\phi^2\big) \, ,
\end{aligned}\ee
where $f(r) = \big(r^2 -2mr -n^2\big)/\big(r^2 +n^2\big)$. Its outer and inner horizons are
determined by $r_{\pm} = m\pm \sqrt{m^2+n^2}$, with their corresponding entropies: $S_{\pm} = \pi
\big(r_{\pm}^2 +n^2\big)$ and Hawking temperatures: $T_{\pm} = (r_{\pm}-m)/[2\pi\big(r_{\pm}^2
+n^2\big)]$. Since the construction uses only thermodynamic quantities at both horizons, it
therefore is independent of any particular macroscopic convention for the Misner string.

The sum and difference of outer and inner horizon entropies gives those of the LM and RM systems,
respectively: \cite{FN1}
\begin{subequations}\begin{align}
S_L &= 2\pi\big(m^2+n^2\big) \, , \label{Ssum} \\
S_R &= 2\pi\, m\sqrt{m^2+n^2} \label{Sdiff}
\end{align}\end{subequations}
with their associated temperatures
\be
T_L = \frac{1}{8\pi\, m} \, , \qquad
T_R = \frac{1}{8\pi\sqrt{m^2+n^2}}
\label{Tsumdiff}
\ee
being determined via Eq. (\ref{hmr}).

We now consider the uniform partition of the mass and the NUT charge to both systems. So the
LM sector is endowed with energy $E_L = m/2$ and  NUT charge $N_L = n/2$. The factors of $1/2$ follow the conventional effective left/right partition \cite{PLB608-251}, while $m$ remains the gravitational mass of the solution. Indeed, differentiating Eq. (\ref{Ssum}) gives
\be
d\Big(\frac{m}{2}\Big) = T_LdS_L +\phi_Ld\Big(\frac{n}{2}\Big) \, ,
\qquad \phi_L = -\frac{n}{m} \, .
\ee
The corresponding Smarr relation,
\be
\frac{m}{2} = 2T_LS_L +\phi_L\frac{n}{2} \, ,
\ee
is also obeyed. Thus, the LM description is charge-like: it closes with the mass and the NUT
charge, with no need for the secondary hair $J_n$. The RM analysis below verifies whether this
charge-like role exhausts the LRM thermodynamics.

The RM description is more intriguing. Considering Eq. (\ref{Sdiff}) either as a squared entropy
formula or as a quadratic mass formula which contains the product $m^2n^2$, there apparently exist
(at least) three minimal admissible possibilities  to treat with it: (I) only $n$ is a variable; (II)
only $J_n = mn$ as an independent variable; (III) both $n$ and $J_n = mn$ are needed simultaneously
to be considered as independent. Here we restrict the diagnostic to these candidates, without claiming to classify all off-shell embeddings of the two-parameter Taub-NUT family.

One can easily show that the former two options can not lead to the correct temperature $T_R$
in Eq. (\ref{Tsumdiff}) and therefore the third option is required among these candidates. For example, if the RM sector
is depicted only by $E_R = m/2$ and $N_R = n/2$, differentiating Eq. (\ref{Sdiff}) then gives: $dS_R
= 2\pi\big[(2m^2+n^2)\, dm +mn\, dn\big]/\sqrt{m^2+n^2}$, which does not reproduce the temperature
$T_R$ in Eq. (\ref{Tsumdiff}). If instead one tries to characterize the RM thermodynamics only
by $E_R = m/2$ and $J_n = mn$, then its entropy after rewritten as: $S_R = 2\pi\sqrt{m^4 +J_n^2}$,
again fails to yield the required RM thermodynamic temperature (\ref{Tsumdiff}).

These failures indicate that neither reduced state space is sufficient for the fixed temperature of
the RM description, motivating a combined response involving both the charge-like variable
$N$ and the rotation-like candidate $J_n = mn$. Neither can be absorbed into the other without spoiling
the temperature fixed by the horizon harmonic mean relations (\ref{hmr}). The test therefore probes
the state space, not just the algebraic possibility of rewriting an entropy formula.

We are therefore left with the third possibility (III) in which the allowed variables ($m, n, J_n$) are homogeneous extensive. They are treated as independent off shell, with $J_n=mn$ imposed only on the physical solution family. Within this restricted class the left/right test is not a change
of notation, because the same variables must reproduce both the RM temperature and the associated Smarr
formula. In this sense, no new independent charge is introduced. It would fail if these relations closed in an admissible homogeneous state space without $J_n$. The squared-mass formula is only an independent check of a homogeneous mass representation.

Within this homogeneous diagnostic state space, the choice
\be
E_R = m/2 \, , \qquad N_R = n/2 \, , \qquad J_n = mn \, .
\ee
passes the simultaneous tests of RM thermodynamic temperature, first law, and Smarr relation.
To see this, let's write
\be
S_R = 2\pi\sqrt{m^4 +(1+w)m^2n^2 -wJ_n^2} \, ,
\label{SR}
\ee
in which the constant $w$ is to be determined.

The squared entropy (\ref{SR}) is a minimal homogeneous form with three independent off-shell variables
($m, n, J_n$) in the enlarged homogeneous state space. We now first fix $w$ by requiring to
reproduce the correct RM thermodynamic temperature (\ref{Tsumdiff}), and then show that both the
first law and Smarr relation hold true simultaneously. Taking independent variations with respect to $(m,n,J_n)$ and then evaluating the coefficients on the physical submanifold $J_n=mn$, we obtain
\be
\frac{dS_R}{2\pi} = \frac{\big[2m^2 +(1+w)n^2\big]\,dm +(1+w)mn\, dn
 -w n\, dJ_n}{\sqrt{m^2+n^2}} \, ,
\ee
and yields the correct RM temperature $T_R = 1/(8\pi\sqrt{m^2+n^2})$ if and only if $w = 1$ for $n\neq0$. With $w=1$, the resulting first law and Smarr relation are
\bea
d\Big(\frac{m}{2}\Big) &=& T_R dS_R +\omega_R\, dJ_n
 +\phi_R\, d\Big(\frac{n}{2}\Big) \, , \\
\frac{m}{2} &=& 2T_R S_R +2\omega_R\, J_n +\phi_R\, \frac{n}{2} \, ,
\eea
where
\be
\omega_R = \frac{n}{4\big(m^2+n^2\big)} \, , \qquad
\phi_R = -\frac{mn}{m^2+n^2} \, .
\ee
In the above, $dJ_n$ is not a pull-back of the differential \cite{FN2}. (See the End Matter also
for the integrability of the LRM's first laws.) The coefficient $\omega_R$ is therefore a right
response coefficient in an enlarged state space, rather than a chemical potential for an independently
conserved charge.

Thus it is the RM description, rather than the macroscopic first law alone, that is sensitive
to the rotation-like thermodynamic secondary hair $J_n = mn$. Within the minimal homogeneous
class considered here, the LRM criterion favors both $N = n$ and $J_n = mn$, supporting a
bi-hair thermodynamic interpretation of the NUT parameter \cite{PRD100-101501}.

\textit{Supporting evidence from hidden conformal symmetry}---As a consistency check, we now
summarize three related observations supporting the compatibility of the bi-hair proposal with hidden conformal symmetry. First, in studying  hidden conformal symmetry, one usually
obtains the following left- and right-temperatures in the rotational $J$-picture directly via defining
the useful conformal coordinates of the Kerr-NUT solution as \cite{JHEP1024078,PRD82-124051,JHEP1010074,
MPLA27-1250046,IJMPA35-2050156,PLB827-136892}
\be
\cT_L = \frac{M^2+N^2}{2\pi\,Ma} \, , \quad
\cT_R = \frac{\sqrt{M^2+N^2-a^2}}{2\pi\, a} \, .
\ee
These hidden-CFT temperatures are distinct from the harmonic-mean temperatures $T_L$ and $T_R$ in Eq.~(\ref{hmr}).
Second, these temperatures can also be related to the horizon first laws through
\be
\delta\, S_{\pm} = \frac{\delta\, E_L}{\cT_L} \pm \frac{\delta\, E_R}{\cT_R} \, ,
\ee
provided that the charges and potentials entering the inner- and outer-horizon first laws are identified consistently. Third, the $J$-picture central charge involves the angular momentum $J=Ma$, consistent with the angular-momentum assignment used in the bi-hair description. In Ref.~\cite{JHEP1024078}, the angular momenta entering the gravitational first law and the CFT $J$-picture are defined differently, so their correspondence requires additional clarification \cite{FN3}.

Using the central charges $c = c_R = c_L = 12Ma$ calculated from
\be
c = 6\frac{\p\mathcal{F}}{\p\, J} = 12J \, , \quad
\mathcal{F} = \frac{S_+S_-}{4\pi^2} = J^2 +M^2N^2 +N^4 \, ,
\ee
the entropies of both horizons can be reproduced microscopically via the Cardy formula as,
\bea
S_{\pm} &=& S_L \pm S_R \equiv \frac{\pi^2}{3}c(\cT_L \pm\, \cT_R)  \nn \\
 &=& 2\pi\, (Mr_{\pm} +N^2) = \pi\big(r_{\pm}^2+a^2+N^2\big) \, .
\label{CardyEntropies}
\eea

Now let's make a comparison of these results from the hidden conformal symmetry with those
ones obtained by extending the above static case to the rotating Kerr-NUT family. It is well-known
that the Bekenstein-Hawking entropies and Hawking temperatures of the outer and inner horizons
($r_{\pm} = M\pm \sqrt{M^2 +N^2 -a^2}$) are \cite{PRD100-101501}
\bea
S_{\pm} = 2\pi\big(Mr_{\pm} +N^2\big) \, , \quad
T_{\pm} = \frac{r_{\pm} -M}{2\pi\big(r_{\pm}^2+a^2+N^2\big)} \, .
\eea
Then exploiting harmonic mean relations (\ref{hmr}), one can easily compute the corresponding
LRM temperatures as
\be
T_L = \frac{1}{8\pi\,M} \, , \quad
T_R = \frac{\sqrt{M^2+N^2-a^2}}{8\pi\big(M^2+N^2\big)} \, ,
\ee
which are related to the CFT left- and right-temperatures in the $J$-picture via
\be
\frac{\cT_L}{T_L} = \frac{\cT_R}{T_R} = \frac{1}{\Omega_R} = 4\frac{M^2 +N^2}{a} \, ,
\ee
where $\Omega_R$ is the angular velocity in the RM description of the rotating parent family.

It should be pointed out that we have treated this comparison only as an external normalization
and left/right organization check, not as the primary definition of the Taub-NUT thermodynamic
state space or as a holographic derivation; no microscopic dual CFT is assumed in deriving
the LRM first laws. The strict non-rotating limit is subtle, since the conventional microscopic
CFT temperatures contain $a$ in the denominator.

The same hidden-conformal structure also displays the rotation-like NUT combination at the entropy
level. The Cardy entropy of the RM system is
\be
S_R = \frac{\pi^2}{3}c\,\cT_R = 2\pi\, M\sqrt{M^2+N^2-a^2} \, ,
\ee
and hence
\be
\frac{S_R^2}{4\pi^2} = M^4 +J_N^2 -J^2 \, , \quad~~ J = Ma \, , \quad J_N = MN \, ,
\ee
where $J_N = MN$ is the rotating counterpart of $J_n = mn$ in the Taub-NUT case.

The above compatibility provides a thermodynamic link between the LRM construction and the
hidden-conformal organization of Kerr-NUT black holes, not an independent derivation of a Taub-NUT
dual CFT or of the secondary hair. The primary evidence for the bi-hair organization nevertheless
comes from the LRM thermodynamic closure established above.

\textit{Conclusions}---We have argued that LRM thermodynamics based on two horizons provides
an additional diagnostic for the thermodynamic state space of spacetimes with NUT charge. Several
macroscopic formulations can each realize their own first law and Smarr relation, so ordinary
macroscopic consistency alone does not uniquely determine the thermodynamic state space
considered here. The left/right decomposition adds information because the LRM temperatures
are fixed via the harmonic mean relations (\ref{hmr}). In the uncharged Taub-NUT case, the
LM description is charge-like and closes with the mass and the NUT charge. The RM first
law, within the restricted homogeneous diagnostic class considered here, closes only when $J_n = mn$
is also included. This is the precise sense in which the LRM evidence favors a bi-hair
thermodynamic organization through $N = n$ and $J_n = mn$.

Our strategy is also consistent with two additional viewpoints. The Christodoulou-Ruffini-type
squared-mass formula supports the homogeneous mass representation, while the hidden conformal
comparison provides an external entropy and temperature normalization check, not a holographic
derivation. In particular, the Cardy-reconstructed RM entropy of the rotating Kerr-NUT solution
can be written in terms of $J = Ma$ and $J_N = MN$, without by itself determining the secondary hair.
The primary diagnostic remains the left/right closure constructed from the two horizons. More
broadly, the thermodynamic structure uncovered here opens a new perspective on the physical nature and
origin of NUT charge \cite{RMP70-427}, pointing toward deeper questions that extend beyond black hole
thermodynamics.

The broader lesson is that the coexistence of two horizons contains thermodynamic information that is
not visible in any single macroscopic first law. The Misner charge (`$\psi-\cN$') formalism and other
formulations provide internally consistent organizations of the macroscopic variables, whereas the
LRM construction probes correlations encoded jointly by the two horizons and reveals which
response channels are required to represent them. In a separate study we have extended the LRM
criterion to Kerr-Newman-NUT spacetimes \cite{2609.33032}, providing further evidence for the bi-hair
organization in the presence of electric charge and angular momentum. Their asymptotically AdS
extensions provide a natural setting for further tests, while similar horizon correlations may help
distinguish competing thermodynamic descriptions \cite{FN4}. Our bi-hair formalism may also apply to the NUT-charged solutions of Kaluza-Klein gravity, Einstein-Maxwell-dilaton-axion theory and
more general STU model in four dimensions. Another challenging task is how to realize the NUT/CFT
correspondence accurately in the spirit of the renowned Kerr/CFT correspondence? Clearly, further work is needed.

As gravitational wave observations, horizon-scale imaging, and tests of Hawking's area law continue to
deepen empirical access to black holes, clarifying the physical content of their thermodynamic variables
becomes increasingly important. In this sense, the LRM construction we have provided turns the
coexistence of two horizons from a geometric feature into a source of additional physical guidance for
black hole thermodynamics.

\textit{Acknowledgments}---This work is supported by the National Natural Science Foundation of China
(NSFC) under Grants No. 12675077, No. 12205243 and No. 12375053, by the China Scholarship Council (CSC), by the Sichuan Science and Technology Program under Grant No. 2026NSFSC0021, and by the Natural Sciences and Engineering Research Council of Canada.

\onecolumngrid

\section*{End Matter}

\twocolumngrid

\textit{Candidate homogeneous embedding and preliminary checks}---To test the hypothesis in
the main text, we use a candidate homogeneous embedding \cite{PRD100-101501,PRD105-124013,
PLB846-138227,2606.03958,PRD108-064034,PRD108-064035} in which the NUT parameter may enter,
besides $M = m$ and $N = n$, through a rotation-like secondary hair $J_n = mn$. This embedding
is not assumed to be the correct state space; it is confronted with the LRM criterion.
For the Taub-NUT spacetime, the entropies and temperatures at both horizons: $r_h\equiv r_{\pm}
= m\pm \sqrt{m^2+n^2}$ are
\bea
S_h &=& \pi\big(r_h^2+n^2\big) = 2\pi\big(mr_h+n^2\big) \, , \\
T_h &=& \frac{r_h-m}{2\pi\big(r_h^2+n^2\big)} = \frac{1}{4\pi\, r_h} \, .
\eea

The first law and Smarr relation take the form \cite{PRD100-101501}
\bea
dM = T_hdS_h +\omega_h dJ_n +\phi_h dN \, , \\
M = 2(T_hS_h +\omega_h J_n) +\phi_h N \, ,
\eea
where
$\omega_h = n/\big(r_h^2 +n^2\big)$, and $\phi_h = -2nr_h/\big(r_h^2 +n^2\big)$.

Here $J_n$ is regarded as a candidate thermodynamic secondary hair, not an extra metric parameter
or independent asymptotic Noether charge. Its independence is off shell: the physical Taub-NUT
family is recovered on $N=n$ and $J_n=mn$, while the partial derivatives defining the potentials
are taken in the enlarged homogeneous state space. The thermodynamic description test asks
whether the fixed RM thermodynamic temperature requires this response channel, not whether
$J_n$ defines an ordinary Gibbs ensemble.

The same candidate variables also admit a Christodoulou-Ruffini-type squared-mass formula
\cite{PRL25-1596,PRD4-3552},
\be
M^2 = \frac{\pi}{S_h}\Big(\frac{S_h}{2\pi} -N^2\Big)^2 +\frac{\pi\, J_n^2}{S_h} \, ,
\ee
which permits the corresponding potentials to be derived by differentiation:
\bea
T_h &=& \frac{\p\, M}{\p\, S_h}\Big|_{J_n,N}
 = \frac{S_h^2 -4\pi^2\big(J_n^2 +N^4\big)}{8\pi\, MS_h^2} \, , \\
\omega_h &=& \frac{\p\, M}{\p\, J_n}\Big|_{S_h,N} = \frac{\pi\, J_n}{MS_h} \, , \\
\phi_h &=& \frac{\p\, M}{\p\, N}\Big|_{S_h,J_n} = -\frac{S_hN -2\pi\, N^3}{MS_h} \, .
\eea

This squared-mass formula is a useful admissibility check: it indicates a homogeneous mass
representation, but the new discriminator is the left/right closure analyzed in the
main text. In addition, the entropy product of the outer and inner horizons reads
\be
S_+S_- = 4\pi^2n^2\big(m^2+n^2\big) = 4\pi^2\big(N^4 +J_n^2\big) \, ,
\ee
which takes a homogeneous form in the extended variables $N$ and $J_n$. Entropy products have
long been used as probes of thermodynamic structure associated with two horizons \cite{PRL106-121301,
PLB807-135521,IJMPA38-2350090}. Here this relation is used only as a preliminary structural check;
it does not select the state space. The actual discriminator is provided by the RM first law
and Smarr relation in the main text.

\textit{Integrability of the LRM first laws}---We can also examine the on-shell consistency of
the LRM differential relations directly on the physical Taub-NUT solution space, without specifying
an off-shell entropy extension. Using
\be
S_R = 2\pi\, m\sqrt{m^2+n^2} \, , \qquad T_R = \frac{1}{8\pi\sqrt{m^2+n^2}} \, ,
\ee
where $T_R$ is uniquely determined by the harmonic mean relation, it is easy to find that for the RM
system, we have
\be
2T_R\, S_R = \frac{m}{2} \, , \label{cftsmarr}
\ee
while
\be
d\Big(\frac{m}{2}\Big) -T_R\, dS_R = \frac{n(n\, dm -m\, dn)}{4\big(m^2+n^2\big)} \, ,
\ee
is integrable only after being multiplied by a factor $F(m/n)/m$ or $F(m/n)/n$, where $F(x)$ is
a function of $x$.

On the other hand, we can use the secondary hair: $J_n = mn$ (so that $n\, dm = dJ_n -m\, dn$)
to rewrite the right hand side of the above equation as:
\bea
\frac{n(dJ_n -2m\, dn)}{4\big(m^2+n^2\big)} &=& \frac{n\, dJ_n}{4\big(m^2+n^2\big)}
 -\frac{mn}{m^2+n^2}\, d\Big(\frac{n}{2}\Big) \nn \\
&=& \omega_RdJ_n +\phi_Rd\Big(\frac{n}{2}\Big) \, ,
\eea
meanwhile we have
\be
2\omega_RJ_n +\phi_R\frac{n}{2} = 0\, ,
\ee
which makes no contribution to the Smarr-like mass formula (\ref{cftsmarr}).

If one would like to replace $J_n = mn$ by $X = mn +\alpha\, n^2$, this shift only
amounts to modifying the NUT potential as $\phi_R\mapsto\phi_R-4\alpha n\omega_R$ for $N_R=n/2$. However, such a replacement lacks the independent geometric motivation identified above for $J_n=mn$.

For the LM system, by virtue of
\be
S_L = 2\pi\big(m^2 +n^2\big) \, , \quad T_L = \frac{1}{8\pi\, m} \, ,
\ee
one can easily obtain
\bea
&& \frac{m}{2} -2T_L\, S_L = -\frac{n^2}{2m}
 = \phi_L\,\Big(\frac{n}{2}\Big) \, , \\
&& d\Big(\frac{m}{2}\Big) -T_L\, dS_L
 = -\frac{n}{2m}\, dn = \phi_L\, d\Big(\frac{n}{2}\Big) \, .
\eea
Thus, the on-shell differential relations reproduce the thermodynamic coefficients obtained
from the enlarged thermodynamic representation, providing a complementary consistency check.

\textit{Representative macroscopic Taub-NUT schemes}---As before, denoting $r_h = r_{\pm}$
for both horizons of the Taub-NUT spacetime, several different formalisms of horizon thermodynamics
can be concisely categorized as follows.

In the \emph{generalized `$\psi-\cN$' formalism} \cite{PRD100-064055,CQG36-194001,PLB798-134972,
JHEP0520084}, one can show that there are infinitely many first laws and Smarr mass formulae:
\bea
dM -T_h\, dS_h &=& \frac{n^2dr_h -3nr_hdn}{2r_h^2} = \psi\, d\cN \, ,\\
M -2T_h\, S_h &=& -\frac{n^2}{r_h} = 2k\psi\cN \, ,
\eea
where
\be
\psi = \frac{r_h^{k-1}}{8\pi n^{3k-2}} \, , \quad
\cN = -\frac{4\pi n^{3k}}{kr_h^k} \, ,
\ee
with $k \ne 0$. The case $k = 1$ recovers the ordinary Misner charge formalism \cite{PRD100-064055,
CQG36-194001}.

In the \emph{modified mass formulation} \cite{PRD101-124011,PRD105-124034}, one usually defines
the internal energy
\be
U = m -n\chi_h = \frac{r_h}{2} \, ,\quad \chi_h = -\frac{n}{2r_h} \, , \quad  N=n \, ,
\ee
so that
\be
dU = T_h \, dS_h +\chi_h dN \, , \quad
U = 2T_h\, S_h +\chi_h N \, .
\ee

The \emph{horizon mass version} \cite{JHEP1022174,PRD100-064055} instead uses the horizon mass
\be
M_h \equiv U -n\chi_h = m -2n\chi_h = m +\frac{n^2}{r_h}
 = \frac{r_h^2 +n^2}{2r_h} \, ,
\ee
to write the mass formulae:
\be
dM_h = T_h \, dS_h +nd\chi_h \, , \quad
M_h = 2T_h\, S_h \, .
\ee
where $\chi_h = -n/(2r_h)$ is viewed as a ``charge".

Finally, in the \emph{$J_n-\Xi$ formalism} \cite{IJMPD31-2250021} with $M = m$, $J_n = mn$, and
$\Xi = nr_h$, the mass formulae read
\bea
dM &=& 2T_hdS_h +\frac{2dJ_n -d\Xi}{2n} \, , \\
M &=& 2T_hS_h +\frac{2J_n -\Xi}{n} \, .
\eea


These mass formulae are not alternatives to be ruled out here; they illustrate that
macroscopic consistency alone is not a unique discriminating principle for the
thermodynamic variables. The LRM criterion instead asks which variables organize the
prescribed thermodynamics of the two horizons.


\begin{thebibliography}{99}
\bibitem{FN1}
In contrast with the Kerr black hole, there is no extremal horizon for the Lorentzian
Taub-NUT case, but the extremal case in the Euclidean case ($n\to i\tilde{n}$) corresponds
to a nut (which has zero entropy) with $m = \tilde{n}$.
\bibitem{FN2}
Remarkably, even without specifying the above off-shell entropy extension, one can directly
start from the RM entropy and temperature on the physical Taub-NUT solution space. Using $J_n = mn$,
the difference between $dE_R$ and $T_RdS_R$ can be expressed in terms of the work contributions
associated with the NUT charge and the secondary hair, yielding the same thermodynamic coefficients
as those obtained from the enlarged thermodynamic representation.
\bibitem{FN3}
In Ref. \cite{JHEP1024078}, the `$\Psi-\cN$' formalism is introduced, and its matching to the CFT $J$-picture requires care: the central charge involves the angular momentum $J = Ma$, whereas the horizon angular momentum also includes the contribution from the Misner string $J_s = n^2a/r_+$. The relation between these angular-momentum definitions should therefore be specified when comparing the gravity and CFT expressions for the left- and right-moving temperatures. Separately, the assumption used to fix the Misner charge in Eq.~(5.3) requires examination in the Taub-NUT limit. In this limit, it leads to $m^2+n^2 = 0$ or $\delta\, m = \delta\, n = 0$, which excludes generic variations required by the first law.
\bibitem{FN4}
For example, different mass prescriptions for four-dimensional Kerr-AdS black holes have been
discussed in Refs. \cite{CQG22-1503,PRD72-044009,2304.10290,2407.09610}. Considering $M_p = m/\Xi^p$,
with $\Xi = 1 -a^2/\ell^2$, the two-horizon consistency test yields $\Xi^{2-p} = 1$, selecting $p = 2$
and hence $M = m/\Xi^2$, corresponding to the nonrotating frame at infinity \cite{CQG22-1503}. Similar
distinctions between rotating and nonrotating frames at infinity also arise in more general rotating
AdS black holes, such as Kerr-Newman-AdS black holes, providing further settings for applying the LRM
criterion.
\end{thebibliography}
\end{document}